\documentclass{optica-article}

\journal{opticajournal} 

\articletype{Research Article}

\begin{document}

\title{A stable phase-locking-free single beam optical lattice with multiple configurations}

\author{Yirong Wang,\authormark{1,2} Xiaoyu Dai,\authormark{1,2} Xue Zhao,\authormark{1,2} Guangren Sun,\authormark{1,2} Kuiyi Gao,\authormark{1,2,*} and Wei Zhang\authormark{1,2,3,$\dagger$}}

\address{\authormark{1}School of Physics and Beijing Key Laboratory of Opto-electronic Functional Materials \& Micro-Nano Devices, Renmin University of China, Beijing 100872, China\\
\authormark{2}Key Laboratory of Quantum State Construction and Manipulation (Ministry of Education), Renmin University of China, Beijing 100872, China\\
\authormark{3}Beijing Academy of Quantum Information Sciences, Beijing 100093, China}

\email{\authormark{*}kgao@ruc.edu.cn,  \authormark{$\dagger$}wzhangl@ruc.edu.cn} 


\begin{abstract*} 
Optical lattices formed by interfering laser beams are widely used to trap and manipulate atoms for quantum simulation, metrology, and computation. To stabilize optical lattices in experiments, it is usually challenging to implement delicate phase-locking systems with complicated optics and electronics to reduce the relative phase fluctuation of multiple laser beams. Here we report a phase-locking-free scheme to implement optical lattices by passing a single laser beam through a prism with $n$-fold symmetric facets and large apex angles. The scheme ensures a stable optical lattice since the interference occurs among different deflected parts of a single laser beam without any moving component. Various lattice configurations, including a triangular lattice and a quasi-crystal lattice with ten-fold symmetry are demonstrated. In both cases, stability measurements show a change of lattice constant in less than $1.14\%$, and a drift of lattice position in less than $1.61\%$.
\end{abstract*}

\section{Introduction}
Atoms trapped in different types of potentials are central subjects under investigation in the field of atomic physics and quantum optics. Due to the exceptional high purity and controllability\cite{NP8}, they have also received extensive attention in quantum simulation\cite{RMP2014}, quantum sensing\cite{Degen2017} and quantum computation\cite{Weiss2017}. The pre-cooled cold atoms are usually trapped by conservative potentials such as magnetic and optical traps. To control the interaction of atoms for quantum simulation, or to isolate them for metrology applications, trapping potentials with different geometric configurations are implemented to prepare bulk gases in optical dipole traps or spatially ordered systems in optical lattices \cite{PRL81}or tweezer arrays. For example, optical lattices, known as spatially repeated optical potentials induced by position-dependent light shift of the inference of coherent laser beams, are widely used to control the spatial positions, the local density and even the interaction of the confined atoms, and trigger a burst of research in simulating important lattice models of various kinds in the last two decades \cite{RMP802008,NRP22020,NP162020}.

The key mechanism of generating a periodic optical potential is using the interference pattern of coherent laser beams to exert periodic conservative forces to the atoms. The configuration of the resulting lattice is determined by the spatial arrangement, the polarization, and the relative phase of the laser beams. For instance, six counter-propagated beams from orthogonal directions can form a simple cubic lattice, in which pioneering experiments of superfluid to Mott insulator transition was first demonstrated \cite{NATURE415}. Beyond that, more complex non-cubic lattice configurations, such as triangular lattice \cite{NJP122010,Yang2021}, honeycomb/hexagonal lattice \cite{N4832012,Uehlinger2013,Jotzu2014}, Kagome lattice \cite{PRL1082012,Leung2020}, superlattice \cite{Folling2007,Lohse2016,Nakajima2016} and even quasi-crystalline lattice \cite{PRL1222019,Yu2024} are also implemented in various experiments to explore the exotic quantum phases of matter.

Meanwhile, the splitting, controlling and overlapping multiple laser beams to form a complex optical lattice usually requires a complicated optical setup and faces some technical difficulties due to its low stability. For example, the geometry of a non-cubic lattice is very sensitive to the relative phases of the laser beams from non-orthogonal directions. A change of phase can deform the geometric configuration and even change the topology of the lattice. To overcome this issue, significant effort needs to be devoted to reduce the relative phase fluctuation by a phase-locking system \cite{Beckerthesis} or even by an additional optical interferometry \cite{Thomasthesis}. Instead of multiple-beam interference in a complicated optical system with a long optical path and delicate feedback control for their phases, multi-facet prism was proposed to create triangular and square lattices\cite{OE58032006}. And a scheme of interfering multiple parallel beams focused by the same lens was reported to generate a dynamic cubic lattice \cite{OE162008}. These approaches naturally lead to more stable lattices since they are mostly composed by simple and fixed optical parts, which can significantly reduce relative phase fluctuation of the beams. So far, this scheme is yet to be implemented for more complex optical lattices.

In this work, we report a new scheme to generate complex non-cubic lattices, such as triangular lattice and quasi-crystalline lattice by shining a single Gaussian beam through a multi-facet azimuthal-symmetric prism. Every portion of the prism acts as a prism and deflects the corresponding part of the Gaussian beam with the same deflecting angle. All deflected parts of the beam meet in the Bessel region and interfere to form a lattice. The relative phases of different parts are determined by their corresponding optical paths before they meet. Since all the interfering beams are from the same Gaussian beam and almost share the same optical path, the consequent relative phases are small and stable. The interference pattern is measured in the Bessel region and also at the atom position by an imaging system with a high-resolution objective. We demonstrate a triangular lattice and a ten-fold quasi-crystalline lattice, which can be obtained and easily switched by just replacing the prism. The lattice patterns and their Fourier transform are analyzed to confirm their periodic characteristics. The time evolution of the interference pattern is recorded to show the stability of the beam, which also includes the pointing stability of the lattice beam itself. From the measurement, the root-mean-squared error of lattice spacing fluctuation is less than $1.14\%$, and a drift of lattice position in less than $1.61\%$ for a measuring time up to 200 minutes. Moreover, the radial envelope of the lattice depth is no-longer harmonic but flat-top like, which helps to reduce effects of inhomogeneous density and phase separation of the trapped atoms. This phase-locking-free single beam optical lattice provides an efficient and compact way to access complex optical lattice for quantum simulation and metrology applications.

\section{Concept and design}

\begin{figure}[htbp]
	\centering\includegraphics[width=12.5cm]{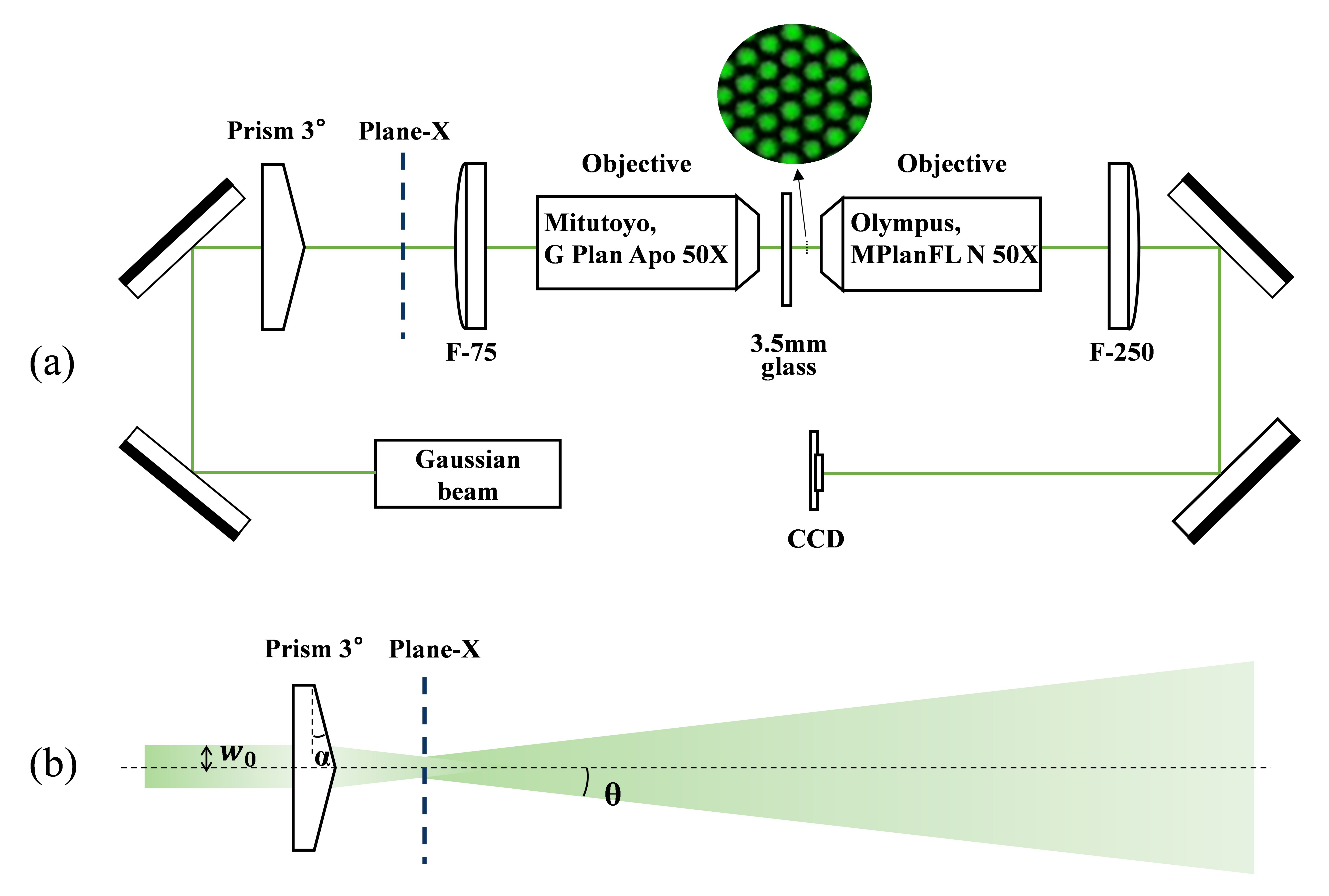}
	\caption{(a) Illustration of setup to generate and picture optical lattices. The lattice potential is realized on plane X by shining a single beams through a prism, and zoomed to a desired size by a telescope ( F-75 lens and Mitutoyo objective). To further visualize the lattices, a second telescope (Olympus objective and F-250 lens) is used. (b) Schematic diagram of the multi-facet prism and light interference.
		\label{fig1}	
	}
\end{figure}

An overview of our phase-locking-free single beam optical lattice scheme is illustrated in Fig. \ref{fig1}(a). A collimated Gaussian beam is shined through a multi-facet prism, and is split into multiple beams. The multi-facet prism acquires an $n$-fold rotational symmetry, with symmetry axis along the optical axis. All beams are deflected toward the optical axis and overlap in the Bessel region, where they interference and form an optical lattice of desired configuration. As shown in Fig. \ref{fig1}(b), the Gaussian beam with electric field 
 ${\vec E}_{0} e^{i\boldsymbol{k}\cdot \boldsymbol{r}}$ incidents from the flat side of the prism, where the wavevector $ \left| k\right|  = 2\pi/\lambda$ and the beam waist is $w_0$. If the transmitted loss is negligible, it is then split into multiple beams with electric field components of ${\vec E}_{0i} e^{i\boldsymbol{k}_i \cdot \boldsymbol{r} + \delta_i}$ with $i = 1, 2, \dots, n$. The wave vector and intensity of each scattered beam are assumed to be identical, i.e., $\left| \boldsymbol{k}_{i} \right| \approx \left| \boldsymbol{k} \right|$ and ${\vec E}_{0i} = {\vec E}_{0}/\sqrt{n}$. The $n$ beams are deflected towards the axis with a small angle $\theta = \left( \mu -1 \right) \alpha $, where $\mu$ is the refractive index of the fused silica prism. Assuming all the $n$ beams have the same polarization, one can calculate the interference pattern after passing through the prism. Ignoring the radial variation of light intensity during light propagation, the distribution of interference light intensity in the radial plane is given by
\begin{eqnarray}
I &=& \frac{\left| E_{0} \right|^{2}}{n} \left[ {\sum\limits_{j = 1}^{n}e^{i{(\boldsymbol{k}_j  \cdot \boldsymbol{r} + \delta_{j}})}}  \right]^2 
\nonumber  \\
&= &\left| E_{0} \right|^{2} + \frac{2\left| E_{0} \right|^{2}}{n} {\sum\limits_{m > j}^{n}{\sum\limits_{j = 1}^{m}{ \cos\left[ \left( k_{\rho}\rho \cos\left( \phi - \phi_{j} \right) - \cos\left( \phi - \phi_{m} \right) \right) + \delta_{j} - \delta_{m} \right]}}},
\end{eqnarray}
where $\rho$ and $k_\rho$ are the spatial coordinate and projected wavevector in the radial plane, respectively, $\phi$ is the azimuthal angle, and $\phi_j = 2\pi (j-1)/n$.
One can easily conclude that the interference pattern also acquires an $n$-fold rotational symmetry. The maximal overlapping area can be reached on plane X, located at a distance of $w_0 / \tan \theta$ from the vertex of the prism. For example, one can obtain a triangular lattice for $n=3$, a cubic lattice for $n=4$, and a ten-fold quasicrystalline lattice for $n=5$. The cases for even higher value of $n$ are not in the scope of this work.

\section{Experiment}

Next we focus on the triangular lattice and ten-fold quasicrystalline lattice, and demonstrate the implementation of our scheme and analyze the lattice stability. For that purpose, the prisms used are the 3- and 5-fold rotationally symmetric prisms. A 532nm laser beam (Coherent Verdi V18, coherent length ~60m) is fiber coupled to the test system and collimated with a beam waist of $w_0 =1.8mm$. Fused silica prisms with a refractive index of $\mu =1.46$ (for 532nm laser) are chosen for high power applications due to their lower lensing effect. The angle of the prisms is $\alpha = 3 ^\circ$, so the deflection angle is $\theta \approx 1.4 ^\circ$.

\begin{figure}[htbp]
	\centering\includegraphics[width=14cm]{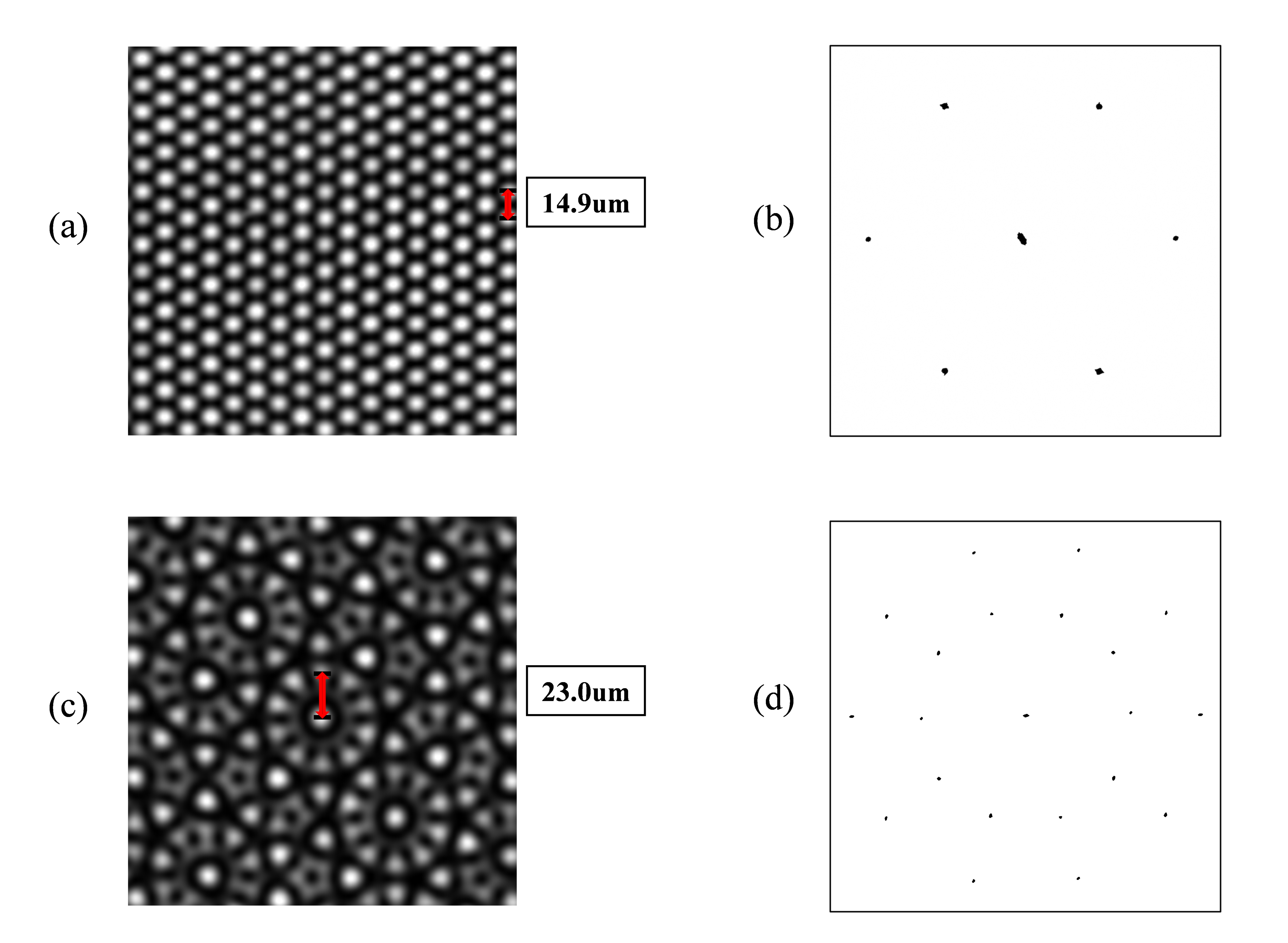}
	\caption{The real-space light-intensity distribution (a) and its Fourier transform (b) of a triangular lattice. The lattice constant (red arrow) is  14.9$\mu$m. The real-space distribution (c) and its Fourier transform (d) of a ten-fold quasi-crystalline lattice. The distance between the two neighboring sites labeled by red arrow is 23.0$\mu$m.}
	\label{fig2}
\end{figure}

A camera (IMAGINGSOURCE, DMK 23UX236) is used to record the light intensity distribution of the interference pattern on plane X. The pixel size of the camera is $2.8\mu m$. By utilizing 3- and 5-fold rotationally symmetric prisms, we obtain a triangular lattice and a ten-fold quasicrystalline lattice, as shown in  Fig. \ref{fig2}. Additionally, a camera with smaller pixel size (Raspberry Pi, module V2) is also used to obtain higher spatial resolution of light intensity distribution for stability measurements. To obtain the lattice constant of the triangular lattice, two typical neighboring lattice sites are randomly chosen. As shown in Fig. \ref{fig2}(a), by Gaussian fitting the light intensity at each site and averaging distances crossing tens of nearest neighboring peaks, we obtain a lattice constant of $14.9 \mu m$, which agrees well with the value of $14.7\mu m$ in the calculated pattern, considering the possible machining precision of this prism. For the ten-fold quasicrystalline lattice demonstrated in Fig. \ref{fig2}(c), the light intensity distribution also consists with the calculated pattern. The distance between the two chosen neighboring lattice sites is $23.0\mu m$. Some defects in the ten-fold quasicrystalline lattice are attributed to the imperfection machining of the prism. 

In order to further confirm the symmetry of the lattices, Fourier transformation of the light intensity distribution of both lattices are implemented. The six-fold rotational symmetric pattern of the Fourier transformed triangular lattice is clearly witnessed in Fig. \ref{fig2}(b). In the case of ten-fold quasi-crystalline lattice, the Fourier transformation shows a ten-fold rotational symmetry pattern in Fig. \ref{fig2}(d), which further proves the rotational symmetry of the quasi-crystalline lattice formed by the interference of five deflected beams after the 5-fold prism. 

Further, we find that the envelop of the intensity distribution of the optical lattices established on plane X is no longer a Gaussian shape, but a flat-top like function. This feature is desirable since it can partially overcome the long-standing problem of inhomogeneity, which is usually inevitable in traditional lattices formed by the interference of counter-propagating Gaussian beams along different directions. In our scheme, taking a Gaussian beam through a double-facet prism as an example for simplicity, the upper half of the cut Gaussian beam has lower intensity at the top and higher intensity at the bottom when it is deflected towards the optical axis. On the opposite, the lower half of the cut beam is stronger at the top and weaker at the bottom. In this way, the two beams compensate each other in their Gaussian intensity, and result in a more uniform overall light intensity distribution. The inhomogeneity can also be reduced by flatening the beam intensity distribution with a spatial light modulator (SLM), but with a much higher cost and technical complexity \cite{Shao2024}.

\subsection{Lattice projection}

To have an optical lattice being useful for quantum simulation in a cold atom experiment, the lattice constant needs to be significantly reduced to sub-micron scale so that atoms can hop between lattice sites and present many-body characteristics. To this aim, as shown in Fig. \ref{fig1}, a telescope with a lens of $f_1=75$mm and a commercial high NA microscope objective (Mitutoyo, G Plan Apo 50X, NA = 0.5, $f_2 = 4$mm) are used to project the lattice potential to the plane where atom are supposed to be. The lattice constant in the atom plane is reduced by a factor of 18.75, leading to a lattice constant of $0.795 \mu$m for the triangular lattice, and a distance of $1.23\mu$m between the two neighboring sites (labeled by red arrow in Fig. \ref{fig2}) for the ten-fold quasicrystalline lattice. We can also estimate the lattice depth from the power and beam waist of the original 532nm Gaussian beam. Taking an experiment with $^6$Li as an example, a Gaussian beam with power of 10W can implement a 100×100 triangular lattice with lattice constant of $0.795 \mu$m and a lattice depth up to $20E_r$, which is high enough to observe many interesting quantum phenomena.

\begin{figure}[htbp]
	\centering\includegraphics[width=12cm]{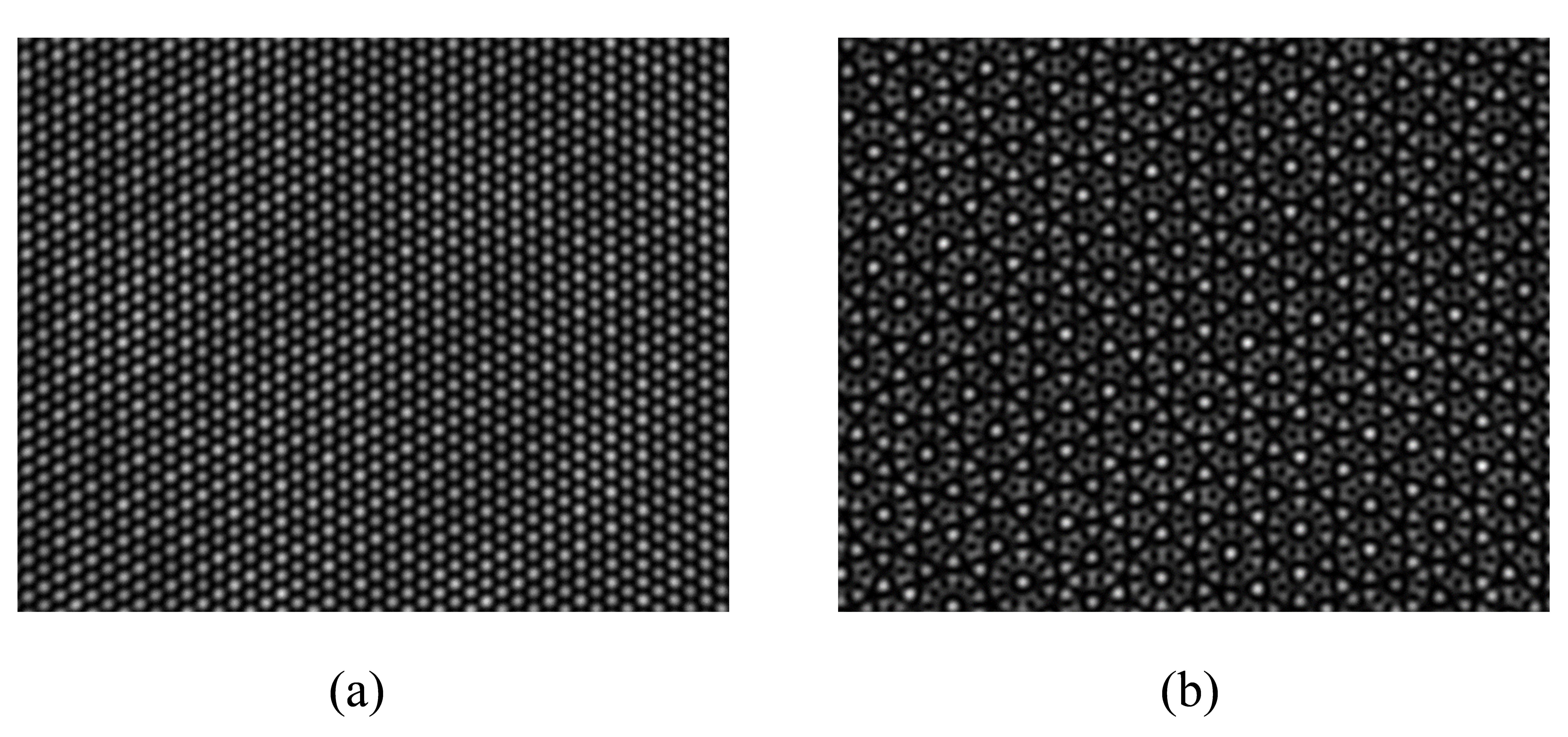}
	\caption{Images of (a) the triangular lattice and (b) the ten-fold quasicrystalline lattice after magnification of the second telescope.}
	\label{fig3}
\end{figure}

To further confirm the detailed structure of the projected optical lattice, we use a second telescope with another microscope objective (Olympus, MPlanFL N 50X, NA = 0.8, $f_3 = 3.6$mm) and a lens with a focal length of $f_4=250$mm. The lattice images are magnified by a factor of 69.4, and are shown in Fig. \ref{fig3}. With the same Gaussian fitting of light intensity distribution as above, a period of $55.7\mu$m is obtained from the magnified image of the triangular lattice, which gives a consistent result of $0.802 \mu$m for the lattice constant before magnification. The first microscope objective is infinity calibrated, so that the $f_1=75$mm lens can be replaced and the lattice constant can be freely chosen for different experimental requirements.

\subsection{Lattice stability}
Non-cubic lattices are usually sensitive to many factors, such as the pointing stability, the relative phase and the polarization of laser beams. In our setup, the total path length is about 200mm, which is about a factor of 10 shorter than that in typical experiments. Additionally, since all different parts of the beam from the multi-facet prism are deflected from the same Gaussian beam, the polarization are almost the same with a small-angled deflection and the relative phases barely change. Those reasons inherently lead to a super-stable lattice without any phase-locking system used in previous experiments \cite{NJP122010,N4832012,PRL1082012}.

\begin{figure}[htbp]
	\centering\includegraphics[width=12cm]{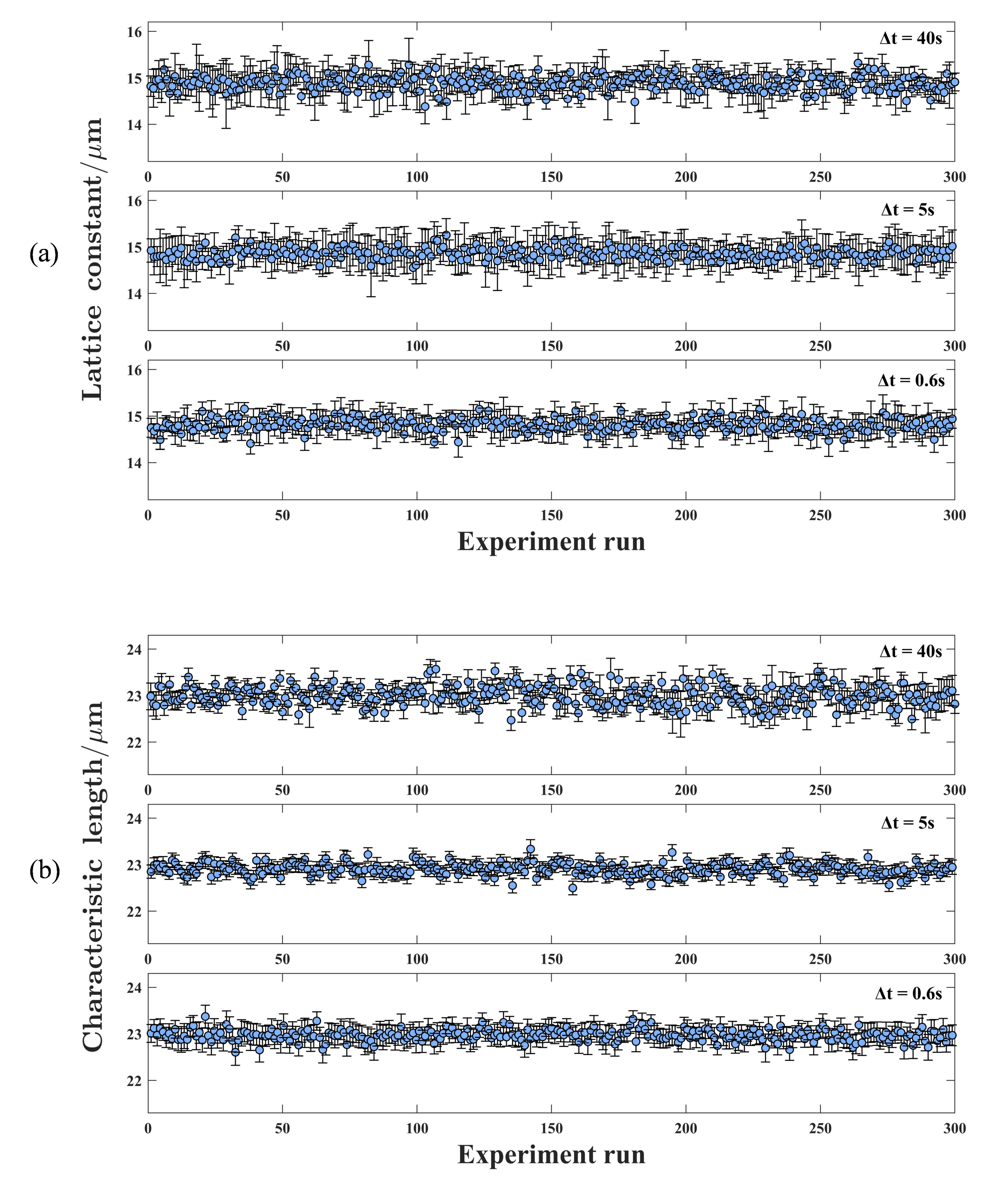}
	\caption{Stability of the lattice potential. (a) Time variation of the lattice constant of the triangular lattice, measured at interval $\Delta t = $40s, 5s, and 0.6s. (b) Time variation of distance between the two neighboring sites of the quasicrystalline lattice. The error bars are calculated from the $95\%$ confidence interval of Gaussian fits when we estimate the center of the interference peaks.}
	\label{fig4}
\end{figure}

To evaluate other possible factors such as mechanical vibration and airflow disturbance that may bring in instability, we use a camera to monitor the variation of the optical lattice over a certain period of time. To eliminate possible problems of the adjustable mounts and translation stages of the objectives, a simple measurement of the lattice stability is taken by monitoring the lattice on plane X behind the prism. The measurement is repeated for every 0.6s, 5s and 40s, respectively. The variation of distances of peaks in the lattice pictures shown in Figs. \ref{fig2}(a) and \ref{fig2}(c) might be attributed to the noise of relative phases and polarizations. Fig.\ref{fig4}(a) shows the variation of lattice constant of the triangular lattice. For short experiment cycles, the root mean square error (RMSE) of the lattice constant is $0.14 \mu$m, while it slightly grows to $0.17\mu$m for long experiment cycles. For quasi-crystalline lattice, the RMSE of the characteristic lattice length for short experiment cycles is less than $0.13\mu$m, and is $0.21\mu$m for long experiment cycles, as shown in Fig. \ref{fig4}(b). The RMSE is less than $1.14\%$ of the lattice constant for the triangular lattice, and is less than $0.91\%$ of the distance between the two neighboring sites for the quasi-crystalline lattice. These results clearly demonstrate the stability of optical lattices obtained by our scheme. We also compare the relative position of the two kinds of lattice on plane X, and the drifts of lattice positions are within $0.24\mu m$ for most of the time. The drift is about $1.61\%$ of the triangular lattice constant, and $1.04\%$ of the quasi-crystalline lattice site distance. 

\section{Summary and outlook}
We experimentally demonstrate a scheme to realize optical lattices of different structures by shining a single beam to multi-facet prisms, and interfering beams diffracted from different facets. Benefits from the simple optical path without any mechanical moving components and phase-locking system, the scheme features ultra-high stability against fluctuations of polarization, optical intensity, and relative phase. For demonstrative purpose, we show a triangular lattice and a ten-fold symmetric quasi-crystalline lattice by using 3-fold and 5-fold symmetric prisms, respectively. In both case, the spacing between lattice sites can be demagnified down to a few hundreds of micrometers by an optical telescope, and the lattice depth can reach $\sim$20$E_r$ with a reasonable laser power of $\sim$10W. To quantify the stability of the optical lattices, we monitor the time variation of lattice spacing, and find a root-mean-squared error of lattice spacing less than $1.14\%$ and a drift of lattice position less than $1.61\%$ for a period of up to 200 minutes. The experimental scheme is very simple, low-cost and easy to be implemented for other configurations and applications. This single-beam lattice provides a super stable lattice potential for further exploration of quantum simulation of exotic phases of quantum matter in complicated lattice system, and might also find more applications in optical tweezers and optical lattice clocks.

\begin{backmatter}
	\bmsection{Funding}
The National Key Research and
	Development Program of China (Grant No. 2022YFA1405301); The National Natural Science Foundation of China (Grants Nos.12274460, 12074428, and 92265208).
\end{backmatter}

\clearpage
\bibliography{wyr0103}

\end{document}